\documentclass{PoS}

\usepackage{graphicx}
\usepackage{dcolumn}
\usepackage{bm}
\usepackage{amssymb}
\usepackage{amsfonts}
\usepackage{amsmath}
\usepackage{latexsym}
\usepackage{dsfont}
\usepackage{multirow}
\usepackage{color}
\usepackage[mathscr]{eucal}

\newcommand{\beq}{\begin{equation}}
\newcommand{\eeq}{\end{equation}}
\newcommand{\beqy}{\begin{eqnarray}}
\newcommand{\eeqy}{\end{eqnarray}}
\newcommand{\beqyn}{\begin{eqnarray*}}
\newcommand{\eeqyn}{\end{eqnarray*}}

\newcommand{\bs}{\begin{slide}}
\newcommand{\es}{\end{slide}}
\newcommand{\bc}{\begin{center}}
\newcommand{\ec}{\end{center}}
\newcommand{\bmin}{\begin{minipage}}
\newcommand{\emin}{\end{minipage}}

\newcommand{\bi}{\begin{itemize}}
\newcommand{\ei}{\end{itemize}}

\newcommand{\bea}{\begin{eqnarray}}
\newcommand{\eea}{\end{eqnarray}}
\newcommand{\be}{\begin{equation}}
\newcommand{\ee}{\end{equation}}

\newcommand{\ud}{\mathrm{d}}

\newcommand{\uTr}{\mathrm{Tr}}

\newcommand{\uD}{\mathcal{D}}

\newcommand{\barpsi}{\overline{\psi}}

\newcommand{\pure}{\text{pure}}
\newcommand{\phys}{\text{phys}}

\newcommand{\LRpartial}{\overset{\leftrightarrow}{\partial}\!\!\!\!\phantom{\partial}}
\newcommand{\LRD}{\overset{\leftrightarrow}{D}\!\!\!\!\!\phantom{D}}

\title{The light-front energy-momentum tensor}

\ShortTitle{The LF EMT}

\author{\speaker{C\'edric Lorc\'e}\\
        Centre de Physique Th\'eorique, \'Ecole polytechnique, CNRS, Universit\'e Paris-Saclay, F-91128 Palaiseau, France\\
SLAC National Accelerator Laboratory, Stanford University, Menlo Park, CA 94025 USA\\
IFPA,  AGO Department, Universit\'e de Li\` ege, Sart-Tilman, 4000 Li\`ege, Belgium\\
E-mail: \email{cedric.lorce@polytechnique.edu}}

\abstract{We present the first complete parametrization for the matrix elements of the generic light-front gauge-invariant energy-momentum tensor, derive the expressions giving separately the spin and orbital angular momentum of quarks and gluons as probed in high-energy scattering experiments, and discuss the relations with two-parton generalized and transverse-momentum dependent distributions. As a by-product, we recovered the Burkardt sum rule, clarified its physical meaning and obtained similar new sum rules for higher-twist distributions.}

\FullConference{QCD Evolution 2015 -QCDEV2015-\\
		26-30 May 2015\\
		Jefferson Lab (JLAB), Newport News Virginia, USA}

\begin{document}

\section{Introduction}

It is well known that the canonical energy-momentum (EMT) obtained from Noether's theorem is usually neither symmetric nor gauge invariant.  In order to cure these ``pathologies'', one often ``improves'' the canonical EMT by adding a so-called superpotential term, \emph{i.e.} a term of the form $\partial_\alpha f^{[\alpha\mu]\cdots}(r)$ where the square brackets stand for antisymmetrization. The physical meaning of this term is a redefinition of the local density of energy and momentum~\cite{Hehl:1976vr,Liu:2015xha} without affecting the total (\emph{i.e.} integrated) linear and angular momentum. Using an appropriate superpotential, Belinfante and Rosenfeld~\cite{Belinfante:1939,Belinfante:1940,Rosenfeld:1940} obtained a new EMT which is both symmetric and gauge invariant. We note in passing that the symmetry requirement for the EMT is essentially motivated by General Relativity where torsion is assumed to vanish. This theory is purely classical and does not incorporate in a consistent manner the quantum concept of spin. In more general theories of gravitation like Einstein-Cartan theory and metric-affine gauge theory, the no-torsion assumption is relaxed leading to asymmetric EMTs and a natural coupling between gravitation and spin. The effects of the latter are however extremely small and are expected to show up only under extreme conditions, see \emph{e.g.}~\cite{Hehl:1976kj,Hehl:1994ue,Obukhov:2014fta} and references therein. 

The early papers about the proton spin decomposition~\cite{Jaffe:1989jz,Ji:1996ek,Shore:1999be} start with the Belinfante-Rosenfeld tensor and introduce additional superpotential terms to decompose the angular momentum into spin and orbital contributions. On the one hand, textbooks like \emph{e.g.}~\cite{Cohen,Simmons} claim that such a decomposition into spin and orbital parts is not possible in a gauge-invariant way for the gauge field. On the other hand, it turns out that the photon spin and orbital angular momentum (OAM) are routinely measured in Quantum ElectroDynamics, see \emph{e.g.}~\cite{Bliokh:2014ara} and references therein. In Quantum ChromoDynamics (QCD), a gauge-invariant quantity called $\Delta G$ has been measured in polarized deep inelastic and proton-proton scatterings, see~\cite{deFlorian:2014yva} for a recent analysis, and can be interpreted in the light-front gauge as the gluon spin~\cite{Jaffe:1989jz}. Motivated by these experimental facts, Chen \emph{et al.}~\cite{Chen:2008ag} claimed in 2008 that the textbooks were wrong and proposed a formal gauge-invariant decomposition of the photon and gluon angular momentum, triggering strong criticism and a multiplication of theoretical papers, summarized in the recent reviews~\cite{Leader:2013jra,Wakamatsu:2014zza}. The dust having settled, it is now understood that the contradiction with the textbook claim is only apparent because the Chen \emph{et al.} construction turned out to be intrinsically non-local~\cite{Hatta:2011zs,Lorce:2012rr,Lorce:2012ce}, whereas textbooks implicitly refered to local quantities only. 

It has actually been known for quite some time that gauge invariance can be restored by allowing the quantities to be non-local~\cite{Dirac:1955uv,Mandelstam:1962mi}. Although there are in principle infinitely many ways of doing this, the experimental setup and the theoretical framework usually determine which is the natural non-local gauge-invariant extension to use~\cite{Lorce:2013bja}. Typical examples of measurable non-local quantities are parton distributions where the gauge invariance is ensured by a Wilson line whose path is determined by the factorization theorems~\cite{Collins:2011zzd}. In particular, it has been shown in Refs.~\cite{Lorce:2011kd,Lorce:2011ni,Hatta:2011ku} that the gauge-invariant form of the canonical OAM is naturally related to so-called Generalized Transverse-Momentum dependent Distributions (GTMDs). These GTMDs are extremely interesting objects since they provide the maximal information about the relativistic phase-space (or Wigner) distribution of quarks and gluons inside the proton. Unfortunately, apart possibly in the low-$x$ regime, it is not known so far how to access these GTMDs experimentally~\cite{Meissner:2009ww}. They are however very useful tools which can be accessed indirectly using realistic models, see \emph{e.g.}~\cite{Lorce:2011kd,Kanazawa:2014nha,Mukherjee:2014nya,Mukherjee:2015aja,Miller:2014vla,Liu:2014vwa,Liu:2015eqa}, or lattice QCD in the infinite-momentum limit~\cite{Musch:2010ka,Musch:2011er,Ji:2013dva,Ji:2014lra,Lin:2014zya,Ma:2014jla}.

There are essentially two families of EMTs in a gauge theory~\cite{Wakamatsu:2010qj,Wakamatsu:2010cb,Lorce:2012rr} : kinetic (or mechanical) and canonical. They all give the same total linear momentum, but attribute different momentum densities to quarks and gluons. The parametrization of the symmetric kinetic (or Belinfante-Rosenfeld) form of the EMT has been given in~\cite{Ji:1996ek} and further discussed in~\cite{Polyakov:2002yz}. The extension to asymmetric kinetic EMTs has been discussed in~\cite{Shore:1999be}, but the correct parametrization in the off-forward case has been given in~\cite{Bakker:2004ib}. Finally, the first complete parametrization of the general EMT with non-locality along the light-front (LF) direction $n$ has been given in~\cite{Lorce:2015lna}.

\section{The gauge-invariant linear and angular momentum tensors}\label{sec2}

Most of the decompositions of the EMT found in the literature can be expressed as combinations of the following five gauge-invariant tensors
\begin{equation}
\begin{aligned}
T^{\mu\nu}_1(r)&=\barpsi(r)\gamma^\mu \tfrac{i}{2}\LRD^\nu\psi(r),\\
T^{\mu\nu}_2(r)&=-2\uTr\!\left[G^{\mu\alpha}(r)G^\nu_{\phantom{\nu}\alpha}(r)\right]+g^{\mu\nu}\,\tfrac{1}{2}\uTr\!\left[G^{\alpha\beta}(r)G_{\alpha\beta}(r)\right]\!,\\
T^{\mu\nu}_3(r)&=-\barpsi(r)\gamma^\mu gA^\nu_\phys(r)\psi(r),\\
T^{\mu\nu}_4(r)&=\tfrac{1}{4}\,\epsilon^{\mu\nu\alpha\beta}\partial_\alpha\!\left[\barpsi(r)\gamma_\beta\gamma_5\psi(r)\right]\!,\\
T^{\mu\nu}_5(r)&=-2\partial_\alpha\uTr\!\left[G^{\mu\alpha}(r)A^\nu_\phys(r)\right]\!,
\end{aligned}
\end{equation}
where $\epsilon_{0123}=+1$ and $\tfrac{i}{2}\LRD^\mu=\tfrac{i}{2}\LRpartial^\mu+gA^\mu$ is the hermitian covariant derivative with $\LRpartial^\mu=\overset{\rightarrow}{\partial}\!\!\!\!\phantom{\partial}^\mu-\overset{\leftarrow}{\partial}\!\!\!\!\phantom{\partial}^\mu$. In particular, $T^{\mu\nu}_1$ and $T^{\mu\nu}_2$ correspond to the kinetic form of the quark and gluon EMTs, respectively, whereas $T^{\mu\nu}_1+T^{\mu\nu}_3$ and $T^{\mu\nu}_2-T^{\mu\nu}_3+T^{\mu\nu}_5$ correspond to their canonical form. The various EMTs can be related to each other~\cite{Leader:2013jra,Lorce:2015lna} using the superpotentials $T^{\mu\nu}_4$ and $T^{\mu\nu}_5$, and the QCD equations of motion
\begin{equation}\label{QCDEOM}
\begin{aligned}
\barpsi(r)\gamma^{[\mu} i\LRD^{\nu]}\psi(r)&=-\epsilon^{\mu\nu\alpha\beta}\partial_\alpha\!\left[\barpsi(r)\gamma_\beta\gamma_5\psi(r)\right]\!,\\
2\!\left[\uD_\alpha G^{\alpha\beta}(r)\right]^c_{\phantom{c}c'}&=-g\,\barpsi_{c'}(r)\gamma^\beta\psi^c(r),
\end{aligned}
\end{equation}
where $c,c'$ are color indices in the fundamental representation and $\mathcal D_\mu=\partial_\mu-ig[A_\mu,\quad]$ is the adjoint covariant derivative. Note that because of the first identity in Eq.~\eqref{QCDEOM}, we can write $T^{\mu\nu}_4(r)=-\tfrac{1}{2}\,T^{[\mu\nu]}_1(r)$ and therefore discard the tensor $T^{\mu\nu}_4(r)$ in the following discussions. 

The gauge-invariant canonical EMT requires the introduction of a pure-gauge field
\begin{equation}
A^\pure_\mu(r)\equiv\tfrac{i}{g}\,\mathcal W(r)\partial_\mu\mathcal W^{-1}(r),
\end{equation}
where $\mathcal W(r)$ (called $U_\pure(r)$ in~\cite{Lorce:2012rr}) is some phase factor which cannot be related in a local way to the original gauge field $A^\mu(r)$ and which transforms as $\mathcal W(r)\mapsto U(r)\mathcal W(r)$ under gauge transformations. The ``physical'' gluon field is then defined as
\begin{equation}
A^\phys_\mu(r)\equiv A_\mu(r)-A^\pure_\mu(r).
\end{equation}
In the gauge where $\mathcal W(r)=\mathds 1$, the gauge-invariant canonical decomposition formally reduces to the Jaffe-Manohar decomposition, and can therefore be considered as a gauge-invariant extension of the latter~\cite{Ji:2012ba,Lorce:2012rr,Lorce:2013gxa,Leader:2013jra}. The phase factor $\mathcal W(r)$ is in principle not unique~\cite{Lorce:2012rr,Leader:2013jra}. Leaving at this stage the phase factor unspecified allows us to consider at once whole \emph{classes} of decompositions differing simply by the precise form of the non-local phase factor.

\section{Parametrization}\label{sec3}

In practice, since we want to relate the matrix elements of the gauge-invariant EMT to measurable parton distributions, we identify the non-local phase factor $\mathcal W(r)$ with a Wilson line $\mathcal W_n(r,r_0)$ connecting a fixed reference point $r_0$ (usually taken at infinity) to the point of interest $r$. According to the factorization theorems~\cite{Collins:2011zzd}, these Wilson lines run essentially in a straight line along the LF direction $n$ to some intermediate point $r_n=r\pm\infty n$, and then in the transverse direction to $r_0$. In some sense, these Wilson lines can be interpreted as the background gluon field generated by the hard part of the scattering. The Wilson line associated with the first part of the path $\mathcal W_n(r,r_n)=\mathcal P\!\left[e^{-ig\int^{\pm\infty}_0n\cdot A(r+\lambda n)\,\ud \lambda}\right] $makes the LF gauge $n\cdot A=0$ special, since in this gauge $\mathcal W_n(r,r_n)=\mathds 1$. The transverse Wilson line $\mathcal W_n(r_n,r_0)$ is associated with the residual gauge freedom and can be set to $\mathds 1$ using appropriate boundary conditions for the gauge field~\cite{Lorce:2012ce,Hatta:2011ku}. 

The matrix elements of the generic LF EMT depend in principle on $n$. More precisely, for a target of mass $M$ they depend on the four-vector $N=\frac{M^2\,n}{P\cdot n}$
which is invariant under rescaling of the lightlike four-vector $n\mapsto \alpha n$. They also depend on the average target momentum $P=(p'+p)/2$, the momentum transfer $\Delta=p'-p$, and the parameter $\eta=\pm 1$ indicating whether the LF Wilson lines are future-pointing ($\eta=+1$) or past-pointing ($\eta=-1$). Since $P\cdot\Delta=0$ and $M^2=P\cdot N=P^2+\Delta^2/4$, the scalar functions parametrizing the generic LF EMT are functions of two scalar variables $\xi=-(\Delta\cdot N)/2(P\cdot N)$ and $t=\Delta^2$. Moreover, they also depend on the parameter $\eta$, and are therefore complex-valued just like the GTMDs~\cite{Meissner:2009ww,Lorce:2013pza}.

Using the techniques from the Appendix A of Ref.~\cite{Meissner:2009ww}, the generic LF EMT for a spin-$1/2$ target can be parametrized as~\cite{Lorce:2015lna}
$\langle p',S'|T^{\mu\nu}_a(0)|p,S\rangle =\overline u(p',S')\Gamma^{\mu\nu}_a(P,\Delta,N;\eta)u(p,S)$ with $a=1,\cdots, 5$ and where $S$ and $S'$ are the initial and final target polarization four-vectors satisfying $p\cdot S=p'\cdot S'=0$ and $S^2=S'^2=-M^2$, and $\Gamma^{\mu\nu}_a$ stands for
\begin{align}
\Gamma^{\mu\nu}_a&=Mg^{\mu\nu}A^a_1+\frac{P^\mu P^\nu}{M}\,A^a_2+\frac{\Delta^\mu\Delta^\nu}{M}\,A^a_3+\frac{P^\mu i\sigma^{\nu\Delta}}{2M}\,A^a_4+\frac{P^\nu i\sigma^{\mu\Delta}}{2M}\, A^a_5\nonumber\\
&+\frac{N^\mu N^\nu}{M}\,B^a_1+\frac{P^\mu N^\nu}{M}\,B^a_2+\frac{P^\nu N^\mu}{M}\,B^a_3+\frac{N^\mu i\sigma^{\nu\Delta}}{2M}\,B^a_4+\frac{N^\nu i\sigma^{\mu\Delta}}{2M}\, B^a_5+\frac{\Delta^\mu i\sigma^{\nu N}}{2M}\,B^a_6+\frac{\Delta^\nu i\sigma^{\mu N}}{2M}\, B^a_7\nonumber\\
&+\left[Mg^{\mu\nu}B^a_8+\frac{P^\mu P^\nu}{M}\,B^a_9+\frac{\Delta^\mu\Delta^\nu}{M}\,B^a_{10}+\frac{N^\mu N^\nu}{M}\,B^a_{11}+\frac{P^\mu N^\nu}{M}\,B^a_{12}+\frac{P^\nu N^\mu}{M}\,B^a_{13}\right]\frac{i\sigma^{N\Delta}}{2M^2}\nonumber\\
&+\frac{P^\mu\Delta^\nu}{M}\,B^a_{14}+\frac{P^\nu \Delta^\mu}{M}\,B^a_{15}+\frac{\Delta^\mu N^\nu}{M}\,B^a_{16}+\frac{\Delta^\nu N^\mu}{M}\,B^a_{17}+\frac{M}{2}\,i\sigma^{\mu\nu}\,B^a_{18}+\frac{\Delta^\nu i\sigma^{\mu\Delta}}{2M}\, B^a_{19}\nonumber\\
&+\frac{P^\mu i\sigma^{\nu N}}{2M}\,B^a_{20}+\frac{P^\nu i\sigma^{\mu N}}{2M}\, B^a_{21}+\frac{N^\mu i\sigma^{\nu N}}{2M}\,B^a_{22}+\frac{N^\nu i\sigma^{\mu N}}{2M}\, B^a_{23}\nonumber\\
&+\left[\frac{P^\mu\Delta^\nu}{M}\,B^a_{24}+\frac{P^\nu \Delta^\mu}{M}\,B^a_{25}+\frac{\Delta^\mu N^\nu}{M}\,B^a_{26}+\frac{\Delta^\nu N^\mu}{M}\,B^a_{27}\right]\frac{i\sigma^{N\Delta}}{2M^2}.\label{param}
\end{align}
For convenience, the notation $i\sigma^{\mu b}\equiv i\sigma^{\mu\alpha}b_\alpha$ has been used and the factors of $i$ have been chosen such that the real part of the scalar functions is $\eta$-even whereas the imaginary part is $\eta$-odd
\begin{equation}
X^a_j(\xi,t;\eta)=X^{e,a}_j(\xi,t)+i\eta\,X^{o,a}_j(\xi,t)
\end{equation}
as a consequence of time-reversal symmetry. The hermiticity constraint implies that the real part of $B^a_j$ with $j\geq 14$ is $\xi$-odd and the imaginary part is $\xi$-even. For the other functions, the real part is $\xi$-even and the imaginary part is $\xi$-odd.

Among all the possible structures allowed by Lorentz and discrete space-time symmetries, only $32$ turned out to be independent, see Appendix A of~\cite{Lorce:2015lna}. Interestingly, this number can alternatively be obtained from the following naive simple counting : the generic EMT $T^{\mu\nu}_a$ has $4\times4=16$ components; the target state polarizations $\pm S$ and $\pm S'$ bring another factor of $2\times 2=4$, but parity symmetry reduces the number of independent polarization configurations by a factor $2$, leading to a total of $32$ independent complex-valued amplitudes. These 32 independent amplitudes are in correspondence with 32 independent Dirac structures, a particular set being given by Eq.~\eqref{param}.

The EMTs $T^{\mu\nu}_1$ and $T^{\mu\nu}_2$ are local and therefore do not depend on $N$ or $\eta$. All the scalar functions must then vanish except the functions $A^{e,a}_j(0,t)$ with $a=1,2$ and $j=1,\cdots,5$. These are linearly related to the standard energy-momentum form factors~\cite{Ji:1996ek,Bakker:2004ib,Leader:2013jra} as follows
\begin{equation}\label{EMFFs}
\begin{aligned}
A_q(t)&=A^{e,1}_2(0,t),&A_G(t)&=A^{e,2}_2(0,t),\\
B_q(t)&=A^{e,1}_4(0,t)+A^{e,1}_5(0,t)-A^{e,1}_2(0,t),&\qquad B_G(t)&=A^{e,2}_4(0,t)+A^{e,2}_5(0,t)-A^{e,2}_2(0,t),\\
C_q(t)&=A^{e,1}_3(0,t),&C_G(t)&=A^{e,2}_3(0,t),\\
\bar C_q(t)&=A^{e,1}_1(0,t)+\tfrac{t}{M^2}\,A^{e,1}_3(0,t),&\bar C_G(t)&=A^{e,2}_1(0,t)+\tfrac{t}{M^2}\,A^{e,2}_3(0,t),\\
D_q(t)&=A^{e,1}_4(0,t)-A^{e,1}_5(0,t),&0&=A^{e,2}_4(0,t)-A^{e,2}_5(0,t).
\end{aligned}
\end{equation}
The first four form factors parametrize the symmetric part of the local gauge-invariant EMT, whereas the last one parametrizes its antisymmetric part.

\section{Linear and angular momentum constraints}\label{sec4}

The parametrization~\eqref{param} is only constrained by space-time symmetries. Conservation of total linear and angular momentum lead to further constraints on the scalar functions. More details about the various additional constraints can be found in~\cite{Lorce:2015lna}.

Contracting the EMT with $\tfrac{1}{2M^2}\,N_\mu$ and considering the forward limit $\Delta\to 0$, gives the average four-momentum in the LF form of dynamics
\begin{equation}\label{momentum}
\langle p^\nu_a\rangle\equiv\frac{1}{2M^2}\,\langle P,S|T^{N\nu}_{a}(0)|P,S\rangle = P^\nu A^{e,a}_2+N^\nu (A^{e,a}_1+B^{e,a}_2)+\delta^{a3}\,\frac{\eta}{2}\,\epsilon^{\nu S}_T\,(B^{o,3}_{18}-B^{o,3}_{20}).
\end{equation}
Interestingly, the last term in Eq.~\eqref{momentum} is naive $\mathsf T$-odd and originates from the potential EMT $T^{\mu\nu}_3$. It can be interpreted as the spin-dependent contribution to the momentum arising from initial and/or final-state interactions, see \emph{e.g.}~\cite{Boer:2015vga} and references therein. Because of the structure $\epsilon^{\nu S}_T\equiv\epsilon^{\nu\mu\alpha\beta}S_\mu n_\alpha \bar n_\beta$ with $\bar n$ the dual lightlike four-vector satisfying $n\cdot\bar n=1$ and such that $P^\mu=(P\cdot n)\bar n^\mu+(P\cdot\bar n)n^\mu$, this naive $\mathsf T$-odd contribution is transverse and requires a transverse target polarization. Since the total four-momentum is $\langle p^\nu\rangle= P^\nu$, we recover from summing over all partons the well-known momentum constraints
\begin{equation}\label{momcons}
\begin{aligned}
\sum_{a=1,2}A^{e,a}_1(0,0)=\sum_{a=q,G}\bar C_a(0)&=0,\\
\sum_{a=1,2}A^{e,a}_2(0,0)=\sum_{a=q,G} A_a(0)&=1.
\end{aligned}
\end{equation}

Having a complete parametrization of the generic LF EMT, we can easily compute the matrix elements of the corresponding OAM tensor $L^{\mu\nu\rho}_{a}(r)=r^\nu T^{\mu\rho}_a(r)-r^\rho T^{\mu\nu}_a(r)$. Because of the explicit factors of position $r$, the matrix elements of the generic LF OAM tensor need to be handled with care~\cite{Bakker:2004ib,Leader:2013jra}. Focusing on the longitudinal component of OAM, we found
\begin{equation}
\langle L^a_L\rangle\equiv\frac{\epsilon_{T\alpha\beta}}{2M^2}\left[i\,\frac{\partial}{\partial\Delta_\alpha}\langle p',S'|T^{N \beta}_{a}(0)|p,S\rangle\right]_{\Delta=0}= \frac{S\cdot N}{M^2}\,A^{e,a}_4(0,0).\label{OAM}
\end{equation}
For a longitudinally polarized target, $S\cdot N=M^2$ and so $A^{e,a}_4(0,0)$ can be interpreted as the average fraction of target longitudinal angular momentum carried by the OAM associated with the EMT $T^{\mu\nu}_a$ in the LF form of dynamics. Similarly, the quark and gluon spin contributions $S^{\mu\nu\rho}_1=\frac{1}{2}\,\epsilon^{\mu\nu\rho\sigma}\,\overline\psi\gamma_\beta\gamma_5\psi$ and $S^{\mu\nu\rho}_2=-2\uTr\!\left[G^{\mu[\nu}A^{\rho]}_\phys\right]$ can be expressed in terms of $L^{\mu\nu\rho}_4$ and $L^{\mu\nu\rho}_5$, respectively. We then found for the longitudinal spin contributions
\begin{equation}\label{spin}
\begin{aligned}
\langle S^q_L\rangle\equiv\frac{1}{2M^2}\,\langle P,S|\tfrac{1}{2}\,\epsilon_{T\alpha\beta}S^{N\alpha\beta}_1(0)|P,S\rangle&=-\frac{1}{2}\left[A^{e,1}_4(0,0)-A^{e,1}_5(0,0)\right]\frac{S\cdot N}{M^2},\\
\langle S^G_L\rangle\equiv\frac{1}{2M^2}\,\langle P,S|\tfrac{1}{2}\,\epsilon_{T\alpha\beta}S^{N\alpha\beta}_2(0)|P,S\rangle&=-\frac{S\cdot N}{M^2}\,A^{e,5}_4(0,0),
\end{aligned}
\end{equation}
where Eq.~\eqref{QCDEOM} has been used to express $L^{\mu\nu\rho}_4$ in terms of $T^{\mu\nu}_1$. The scalars $-\tfrac{1}{2}[A^{e,1}_4(0,0)-A^{e,1}_5(0,0)]=-\tfrac{1}{2}D_q(0)$ and  $-A^{e,5}_4(0,0)$ can therefore be interpreted as the average fraction of target longitudinal angular momentum carried by the spin of quarks and gluons, respectively. Adding the spin and OAM contributions, we naturally recover the Ji relation for total angular momentum~\cite{Ji:1996ek}
\begin{equation}\label{AMtot}
\begin{aligned}
\langle J^q_L\rangle&=\langle S^q_L\rangle+\langle L^q_L\rangle=\tfrac{1}{2}\left[A^{e,1}_4(0,0)+A^{e,1}_5(0,0)\right]\tfrac{S\cdot N}{M^2}=\tfrac{1}{2}\left[A_q(0)+B_q(0)\right]\tfrac{S\cdot N}{M^2},\\
\langle J^G_L\rangle&=\langle S^G_L\rangle+\langle L^G_L\rangle=\tfrac{1}{2}\left[A^{e,2}_4(0,0)+A^{e,2}_5(0,0)\right]\tfrac{S\cdot N}{M^2}=\tfrac{1}{2}\left[A_G(0)+B_G(0)\right]\tfrac{S\cdot N}{M^2}.
\end{aligned}
\end{equation}
Finally, since the total angular momentum is $1/2$, we naturally recover from Eqs.~\eqref{momcons} and~\eqref{AMtot} the angular momentum constraint
\begin{equation}\label{cons3}
\sum_{a=1,2}[A^{e,a}_4(0,0)+A^{e,a}_5(0,0)-A^{e,a}_2(0,0)]=\sum_{a=q,G}B_a(0)=0
\end{equation}
also known as the anomalous gravitomagnetic moment sum rule~\cite{Teryaev:1999su,Brodsky:2000ii}.

\section{Link with measurable parton distributions}\label{sec5}

The scalar functions parametrizing the generic LF EMT can be related to measurable parton dsitributions, like \emph{e.g.} Generalized Parton Distributions (GPDs) accessed in exclusive scatterings~\cite{Diehl:2003ny} and Transverse-Momentum dependent Distributions (TMDs) accessed in semi-inclusive scatterings~\cite{Collins:2011zzd}. Both kinds of distributions can be seen as particular projections of the GTMD correlator~\cite{Meissner:2009ww,Lorce:2011dv,Lorce:2013pza}
\begin{equation}
\begin{aligned}
F^\mu_{S'S}(P,x,\Delta,N)&=\int\ud^2k_\perp\,W^\mu_{S'S}(P,x,\vec k_\perp,\Delta,N;\eta),\\
\Phi^\mu_{S'S}(P,x,\vec k_\perp,N;\eta)&=W^\mu_{S'S}(P,x,\vec k_\perp,0,N;\eta).
\end{aligned}
\end{equation}
The matrix elements of the EMT we are interested in can also easily be expressed in terms of these GTMDs~\cite{Lorce:2012ce} and hence related to GPD and TMD correlators
\begin{equation}
\langle p',S'|T^{\mu\nu}(0)|p,S\rangle=\int\ud^4k\,k^\nu\,W^\mu_{S'S}.
\end{equation}
The detailed relations between the EMT scalar functions and two-parton GPDs and TMDs of any twist can be found in~\cite{Lorce:2015lna}.

Among the interesting results, let us just mention that we naturally recover the Burkardt sum rule~\cite{Burkardt:2003yg,Burkardt:2004ur} 
\begin{equation}
\sum_{a=q,G}\int\ud x\,\ud^2k_T\,\tfrac{k^2_T}{2M^2}\,f^{\perp a}_{1T}(x,k^2_T)=0,
\end{equation}
and derived three new similar sum rules for high-twist distributions
\begin{equation}\label{newSR}
\begin{split}
\sum_{a=q,G}\int\ud x\,\ud^2k_T\,\tfrac{k^2_T}{2M^2}\,f^{\perp a}(x,k^2_T)&=0,\\
\sum_{a=q,G}\int\ud x\,\ud^2k_T\,\tfrac{k^2_T}{2M^2}\,f^{\perp a}_L(x,k^2_T)&=0,\\
\sum_{a=q,G}\int\ud x\,\ud^2k_T\,\tfrac{k^2_T}{2M^2}\,f^{\perp a}_{3T}(x,k^2_T)&=0,
\end{split}
\end{equation}
all of them expressing the fact that the total transverse momentum (w.r.t. the target momentum) has to vanish. Higher-twist TMDs are much harder to test experimentally, but it would be very interesting to test these new sum rules using phenomenological models, Lattice QCD and perturbative QCD.

\section{Conclusions}\label{sec6}

A gauge-invariant canonical energy-momentum tensor can be defined once one relaxes the assumption of strict locality without harming causality. This indicates that the canonical energy-momentum tensor \emph{can} be considered as a physical object and \emph{a priori} measured experimentally \emph{via} particular moments of parton distributions extracted from numerous physical processes.

We presented here the complete parametrization for the matrix elements of the generic light-front gauge-invariant energy-momentum tensor and discussed the constraints of linear and angular momentum conservation. We showed that this energy-momentum tensor can be related to moments of the parton distributions in momentum space. Among the interesting results, we recovered the Burkardt sum rule and derived three new sum rules involving higher-twist distributions, all expressing basically the conservation of transverse momentum. We expect highly valuable insights into these matters in a near future coming from new experimental data obtained in existing and future facilities, and explicit investigations using covariant models, Lattice QCD and perturbative QCD.

\acknowledgments

I am thankful to S. Brodsky, E.~Leader, B.~Pasquini and P. Schweitzer for useful discussions related to this study. This work was supported by the Belgian Fund F.R.S.-FNRS \emph{via} the contract of Charg\'e de recherches.

\end{document}